\documentclass[10pt, a4paper]{llncs}

\usepackage{cite}
\usepackage{amssymb, amsmath}
\usepackage{epsfig}
\usepackage{algorithm}
\usepackage[noend]{algpseudocode}

\DeclareMathOperator{\Mat}{Mat}

\DeclareMathOperator{\Nm}{Nm}

\DeclareMathOperator{\vect}{vec}

\DeclareMathOperator{\Gal}{Gal}

\DeclareMathOperator*{\argmin}{arg\,min}
\DeclareMathOperator*{\argmax}{arg\,max}

\DeclareMathOperator{\Cl}{Cl}

\newcommand{\Z}{\mathbb{Z}}
\newcommand{\Q}{\mathbb{Q}}

\newcommand{\C}{\mathbb{C}}
\newcommand{\K}{K}
\newcommand{\LL}{L}

\newcommand{\LK}{\LL/\K}
\newcommand{\GLK}{\text{Gal}(\LK)}
\newcommand{\OK}{\mathcal{O}_{\K}}
\newcommand{\OL}{\mathcal{O}_{\LL}}

\def\lskip{\par \vskip\baselineskip}

\begin{document}
	
\title{Reduced Complexity Decoding of $n\times n$ Algebraic Space--Time Codes} 
\author{Amaro Barreal \and Camilla Hollanti \and David Karpuk \thanks{The authors are financially supported by Academy of Finland grants \#276031, \#282938, \#283262 and \#268364, and a grant from Magnus Ehrnrooth Foundation, Finland. The support from the European Science Foundation under the COST Action IC1104 is also gratefully acknowledged.}} 
\institute{Aalto University \\ Department of Mathematics and Systems Analysis \\ \email{firstname.lastname@aalto.fi}} 

\maketitle

\begin{abstract}
	Algebraic space--time coding allows for reliable data exchange across fading  multiple-input multiple-output channels. A powerful technique for decoding space--time codes is Maximum-Likelihood (ML) decoding, but well-performing and widely-used codes such as the Golden code often suffer from high ML-decoding complexity.  In this article, a recursive algorithm for decoding general algebraic space--time codes of arbitrary dimension is proposed, which reduces the worst-case decoding complexity from $O(|\mathcal{S}|^{n^2})$ to $O(|\mathcal{S}|^n)$.
\end{abstract}

\section{Introduction}

Algebraic Space--Time (ST) coding is a powerful technique for reliable data exchange across fading Multiple-Input Multiple-Output (MIMO) channels, which makes use of available multiple antennas at the transmitter and receiver, as well as multiple channel uses to introduce redundancy in the transmitted information. Data transmission across a MIMO channel can be modeled as
\begin{equation}
\label{eqn:mimo}
	Y_{n_r\times T} = H_{n_r \times n_t}X_{n_t \times T} + N_{n_r \times T},
\end{equation}
where the subscripts $n_t$, $n_r$ and $T$ denote the number of antennas at the transmitter, receiver, and the coding delay, respectively, $Y$ and $X$ are the received and sent matrices, $H$ is a random complex matrix modeling Rayleigh fading, and $N$ is a noise matrix whose entries are complex Gaussian with zero mean.  In this paper we restrict ourselves to the fully symmetric case where $n_t = n_r = T = n$ for some $n$, and thus omit the above subscripts.

First introduced two decades ago, ST codes have been the subject of extensive study, and various algebraic criteria have been derived for ensuring reliable performance. The complexity of the decoding algorithm is a major issue in practical implementation.  In digital video broadcasting \cite{dvb}, high decoding complexity prevents the use of high-rate space-time codes, since the complexity of linear decoders such as the sphere decoder grows exponentially in rank.

For the remainder of this article, we let $\vect: \Mat(m\times n, \C) \to \C^{mn}$ be the vectorization transformation and we denote by $\mathcal{S}$ the signaling alphabet (\emph{e.g.} $4$-QAM) used. Write $\vect(HX) = Gs$, where $s \in \mathcal{S}^k$. Decoding amounts to solving
\begin{equation}
	\argmin\limits_{X \in \mathcal{X}}{\left|\left|Y-HX\right|\right|_F^2} \leadsto \argmin\limits_{s \in \mathcal{S}^k}{\left|\left|\vect(Y) - Gs\right|\right|^2},
\end{equation}
where $||\cdot||_F$ and $||\cdot||$ denote the Frobenius and Euclidean norm, respectively. The latter search can be performed with the help of a sphere-decoder \cite{viterbo}, whose decoding complexity is upper-bounded by $|\mathcal{S}|^k$. 

Cleverly constructed algebraic ST codes known as \emph{fast-decodable} codes reduce the complexity of ML-decoding by parallelizing the ML search (see \emph{e.g.} \cite{biglieri}).  However, ST codes exist that, although they do not allow for a reduced ML-decoding complexity, exhibit other desirable properties. An example of such a code is the Golden code \cite{belfiore}, incorporated in the IEEE 802.16 (WiMAX) standard. The main algorithm of \cite{sirinaunpiboon} reduces the decoding complexity of the Golden code from $O(|\mathcal{S}|^3)$, corresponding to the complexity of ML-decoding, to $O(|\mathcal{S}|^2)$, while maintaining nearly-ML performance. The algorithm is specific to the Golden code, but has been generalized to the $3\times 3$ and $4\times 4$ perfect codes in, respectively, \cite{howard} and \cite{amani}.  The main contribution of this article is a recursive algorithm for decoding arbitrary $n\times n$ algebraic ST codes generalizing those of \cite{sirinaunpiboon,howard,amani}, which reduces the worst-case decoding complexity of $O(|\mathcal{S}|^{n^2})$ down to $O(|\mathcal{S}|^{n})$.
	
We review how algebraic ST codes are constructed from central simple algebras in Section~\ref{sec:stc}. Our algorithm for decoding general $n\times n$ algebraic ST codes is illustrated in Section~\ref{sec:nxn}, wherein we also discuss its complexity.  We then briefly consider the case $n = 2$ and the Golden code as a special example in Section~\ref{sec:2x2}.  In Section~\ref{sec:sims}, we study the $4\times 4$ perfect code in more detail and provide an elementary statistical analysis of the involved quantities.  While preliminary simulations seem to indicate that our algorithm has performance which is close to ML-decoding, time and space constraints prevent us from providing proper error rate simulations in this current draft.  Detailed error rate simulation results will appear in a subsequent version of the paper.

\section{Space--Time Codes from Central Simple Algebras}
\label{sec:stc}

Let $\LK$ be a Galois number field extension of degree $n$, and let $\mathcal{A}$ be a $\K$-central simple $\LL$-algebra of rank $n$ over $\LL$, which we treat as a \emph{right} $L$-vector space.  For the rest of the article, we fix compatible embeddings of $K$ and $L$ into $\C$, and identify $K$ and $L$ with their images under these embeddings.  To construct matrices to serve as codewords, we consider an injective algebra homomorphism
\begin{equation}
	\rho: \mathcal{A} \to \Mat(n,\C). 
\end{equation}
\begin{definition}
An \emph{algebraic space--time code} is a finite subset $\mathcal{X}$ of $\rho(\mathcal{A})$ or its transpose $\rho(\mathcal{A})^t$.
\end{definition}
As a general reference for applications of central simple algebras to space-time coding, we recommend \cite{CDA_green_book}, and as a catch-all mathematical reference for central simple algebras, we will use \cite[Chapter IV]{milne_CFT}.  There are many properties that can be desirable for an algebraic ST code, such as full-rank codewords, non-vanishing determinants, fast-decodability, balanced energy across transmitters, and optimal diversity-multiplexing tradeoff, which can be achieved by demanding some additional algebraic requirements of the involved structures.

Let us fix a basis $\{e_1,\ldots,e_n\}$ of $\mathcal{A}$ as a right $L$-vector space.  As an example of the above, consider the \emph{left regular representation} (LRR), defined as follows.  For fixed $x\in \mathcal{A}$, the map given by left multiplication $y\mapsto xy$ for all $y \in \mathcal{A}$ is right $L$-linear and compatible with algebra multiplication, and thus allows us to represent $x$ as a complex matrix $\rho(x)$ in the above basis.  For notational convenience, we will below mostly consider the transpose of this representation.

To give a concrete example, consider the case where $L/K$ is a cyclic extension, whose Galois group $\text{Gal}(L/K) = \langle \sigma \rangle$ is generated by a single element.  Fix some $\gamma\in K^\times$ and consider the central simple algebra
\begin{equation}
\mathcal{A} = \bigoplus_{j = 0}^{n-1}e_j L
\end{equation}
where the multiplication is determined by
\begin{align}
\lambda e_j &= e_j\sigma^j(\lambda)\quad \text{for all $\lambda\in L$, and} \\
e_je_k &= \left\{\begin{array}{rll}
e_{j+k}\ \ & \text{if}\ \ & j+k \leq n-1 \\
e_{j+k-n}\gamma\ \ & \text{if}\ \ & j+k > n-1
\end{array}\right.
\end{align}
If $\rho$ denotes the LRR, the explicit structure of the codewords of $\mathcal{X}\subset \rho(\mathcal{A})^t$ can be found in, for example, \cite{CDA_green_book}.  More generally one can consider the left regular representation of the crossed-product algebra $\mathcal{A}(\varphi)$ for any $2$-cocycle $\varphi:\Gal(L/K)^2\rightarrow L^\times$ \cite[Chapter IV]{milne_CFT}.

% Denote by $\OK$ and $\OL$ the rings of integers of $\K$ and $\LL$, respectively.  From now on we assume that $K$ has class number $\Cl_{\K} = 1$, which guarantees we can choose a basis $\left\{\omega_1,\ldots,\omega_n\right\}$ of $\OL$ as an $\OK$-module
% \begin{definition}
% 	Let $x = \sum_{i=1}^{n}{e_i x_i} \in \mathcal{A}$ and $\alpha \in \LL^{\times}$ a \emph{shaping element}. A \emph{space--time code} is a subset 
% \begin{equation}
% 	\mathcal{X} \underset{\text{\tiny{finite}}}{\subset} \Ima(\rho(\alpha\mathcal{A})) = \left\{\left. \rho(\alpha)\rho(x) = \rho(\alpha)\sum\limits_{i=1}^{n}{\rho(\omega_i)\rho(x_i)} \right| x \in \mathcal{A}\right\}.
% \end{equation} 
% \end{definition}

%We briefly recall the most important design criteria for ensuring a reliable performance. Let $X,X'$ denote distinct code matrices ranging over a code $\mathcal{X}$.
%\begin{enumerate}
%	\item \emph{Diversity gain:} $\min_{X \neq X'} \rk(X-X') = \min\left\{n_t,T\right\}$. A ST code satisfying this criterion is called a \emph{full-diversity} code.

%	\item \emph{Coding gain:} $\Delta_{\min} := \min_{X \neq X'} \det[(X-X')(X-X')^{\dagger}]$ should be (after normalization to unit volume) as big as possible. If $\Delta_{\min}$ does not vanish as the code size goes to $\infty$, the ST code is said to have the \emph{nonvanishing determinant} property. 
%\end{enumerate} 

Let us return to the general case of $L/K$ a Galois extension of degree $n$, and $\mathcal{A}$ any $K$-central simple $L$-algebra of rank $n$ over $L$.  Consider any injective algebra homomorphism $\rho:\mathcal{A}\rightarrow \text{Mat}(n,\C)$.  The following decomposition of the matrix $\rho(x)^t$ for $x\in \mathcal{A}$ and its vectorization are central to our decoding algorithm.

Given $x\in\mathcal{A}$, we can represent it as $x = \sum_{j = 1}^n e_jx_j$ for $x_j\in L$.  If we fix $\{\omega_1,\ldots,\omega_n\}$ to be a basis for $L$ as a $K$-vector space, then we can write $x$ as
\begin{equation}
x = \sum_{j = 1}^ne_jx_j = \sum_{j = 1}^n e_j \left(\sum_{i = 1}^n\omega_i x_{ji}\right) = \sum_{i = 1}^n \sum_{j = 1}^n e_j\omega_i x_{ji},\quad \text{where $x_{ji}\in K$}
\end{equation}
Passing to the transpose of the representation $\rho$ gives
\begin{equation}
\rho(x)^t = \sum_{i = 1}^n\sum_{j = 1}^n\rho(e_j\omega_i x_{ji})^t 
% = \sum_{i = 1}^n\sum_{j = 1}^n\rho(x_{ji})\rho(\omega_i)^t\rho(e_j)^t 
= \sum_{i = 1}^n\rho(\omega_i)^t\left(\sum_{j = 1}^n\rho(e_j)^t\rho(x_{ji})\right)
\end{equation}
where we have used the fact that $\rho(x_{ji})$ is a scalar matrix and thus commutes with everything.  If we define $X_i = \sum_{j = 1}^n\rho(e_j)^t\rho(x_{ji})$, then vectorizing $\rho(x)^t$ gives, by basic properties of the vectorization operator,
\begin{equation}
\boxed{\vect(\rho(x)^t) = \sum_{i = 1}^n (I_n\otimes \rho(\omega_i)^t)\vect(X_i) = \sum_{i = 1}^n (I_n\otimes \rho(\omega_i)^t)\Gamma_is_i}
\end{equation}
where $s_i = [x_{1i}\cdots x_{ni}]^t$ and $\Gamma_i\in\text{Mat}(n^2\times n,\C)$ is the unique matrix satisfying $\Gamma_i = [\Gamma_{1i}^t \cdots  \Gamma_{ni}^t]^t$ where $\Gamma_{ji}s_i$ is the $j^{th}$ column of $X_i$.  Note that the above also provides us with a decomposition of the generator matrix $G$ of the code, as the block matrix
\begin{equation}
G = [(I_n\otimes \rho(\omega_1)^t)\Gamma_1\cdots (I_n\otimes \rho(\omega_n)^t)\Gamma_n]
\end{equation}
It is worth pointing out that in the case of the cyclic algebra constructed in the above example, the matrices $\Gamma_i$ are the same for all $i$.

The elements $x_{ji}\in K$ are the information symbols we wish to decode, and one should think of them as belonging to our signaling alphabet $\mathcal{S}$.  To introduce some discrete structure, one usually assumes that $x_{ji}\in\mathcal{O}_K$, where $\mathcal{O}_K$ denotes the ring of integers of $K$.  The basis $\omega_1,\ldots,\omega_n$ is often then chosen to be a basis for $\mathcal{O}_L$ as an $\mathcal{O}_K$-module (assuming this is possible), or more generally, a basis of an ideal $\mathcal{I}\subseteq \mathcal{O}_L$ as an $\mathcal{O}_K$-module.  In the case where $\mathcal{I} = (\alpha)\subseteq\mathcal{O}_L$ is a principal ideal, the effect of considering this ideal amounts to replacing $\omega_i$ with $\omega_i\alpha$ in the above expressions.  Such an $\alpha$ is often called a \emph{shaping element}.

\section{The General $n\times n$ Case}
\label{sec:nxn}

In this section we present the main result of the paper, which is a recursive decoding algorithm for $n\times n$ algebraic ST codes with complexity $O(|\mathcal{S}|^n)$.  Informally, the algorithm uses the decomposition of $\vect(X)$ of Section~\ref{sec:stc} to split up the $n^2$ information symbols contained in an $n\times n$ codeword into $n$ vectors, each containing $n$ symbols. Based on how well-behaved certain channel sub-matrices are, we then estimate one of these vectors, requiring a search over the space $|\mathcal{S}|^n$.  This estimate is subtracted from the received signal and we then proceed by recursion.  

Our algorithm performs $n$ searches over the space $|\mathcal{S}|^n$, rather than the single search over the space $|\mathcal{S}|^{n^2}$ required by the naive brute-force algorithm.  Hence it is clear that the complexity is $O(|\mathcal{S}|^n)$, when measured in terms of the size of the space we must search over when solve the below ``$\argmin$''-type problem.  However, one pays for the reduction in complexity in this operation by computing several covariance matrices, evaluating their determinants and condition numbers, and applying a lattice quantizer.  Hence, as with the main algorithm of \cite{sirinaunpiboon} which we are generalizing, there is a non-trivial amount of preprocessing to be done.  This algorithm also generalizes the main algorithm of \cite{amani}, wherein the authors break up the $4\times 4$ perfect code into two groups of eight symbols, as well as the approach in \cite{howard} which studies the $3\times 3$ perfect code.

Let $\LK$, $\mathcal{A}$, and $\rho$ be now as in Section~\ref{sec:stc}, and consider an algebraic ST code of the form $\mathcal{X} \underset{\text{\tiny{finite}}}{\subset} \rho(\mathcal{A})^t$.  Given the channel equation (\ref{eqn:mimo}) for a channel matrix $H = (h_{kl})_{1 \le k,l \le n}$, we first vectorize the expression to get $\vect(Y) = \vect(HX) + \vect(N).$  By similar arguments as in Section~\ref{sec:stc}, we have
\begin{align}
	\vect(HX) = \sum\limits_{i=1}^{n}{(I_n\otimes (H\rho( \omega_i)^t))\Gamma_is_i} = \sum\limits_{i=1}^{n}{H_i s_i},
\end{align} 
where, using the same notation as in Section~\ref{sec:stc}, $s_i = [x_{1i}\ \cdots\ x_{ni}]^{t}$ is the vector containing $n$ of the $n^2$ symbols we wish to decode, and $H_i = (I_n\otimes (H\rho(\omega_i)^t))\Gamma_i \in \Mat(n^2\times n, \C)$.  Setting $y := \vect(Y)$, we can rewrite equation (\ref{eqn:mimo}) as $y = \sum\limits_{i=1}^{n}{H_i s_i} + \vect(N)$.  Decoding the vector $y$ now amounts to solving
\begin{equation}
\label{eqn:argmin}
	\argmin\limits_{(s_1,\ldots,s_n) \in (\mathcal{S}^n)^n}{\left|\left|y-\sum\limits_{i=1}^{n}{H_i s_i}\right|\right|^2}.
\end{equation}
The idea of the decoding algorithm is to first decode some $s_{i_1}$, use this information to recursively decode some $s_{i_2}$, and so on until ultimately decoding $s_{i_n}$ in the $n^{\text{th}}$ step. Define $G_i = \begin{bmatrix}H_1 & \cdots & \widehat{H}_i & \cdots & H_n\end{bmatrix}$.  Let $\kappa(\cdot)$ denote the condition number of any matrix.  The algorithm chooses which vector of symbols to decode first based on the following decision. 
\begin{equation}
\boxed{
\text{Decode $s_{i_1}$ first, where $i_1 = \argmax\limits_{i \in \left\{1,\ldots,n\right\}}{\left\{\frac{\det(G_i^{\dagger}G_i)}{\kappa(G_i^{\dagger}G_i)}\right\}}$}
}
\end{equation}

As is noted in \cite{sirinaunpiboon}, the determinant of the above covariance matrix measures the instantaneous SNR of the corresponding linear system and thus should be large, and the condition number measures the accuracy zero-forcing approximations and thus should be small.  To decode $s_{i_1}$ we will express the other symbols as a function of $s_{i_1}$, and for this function to be as tolerant to interference and noise as possible, one should maximize the above determinant and minimize the above condition number.  Thus we suggest maximizing their ratio as a compromise.

Reindex the terms $H_i s_i$ such that $i_1 = 1$ and define $t_1 = \begin{bmatrix}s_2^t & \cdots & s_n^t \end{bmatrix}^t$. Then, the expression involved in equation (\ref{eqn:argmin}) can be rewritten as 
\begin{align}
	\left|\left|y-\sum\limits_{i=1}^{n}{H_i s_i}\right|\right|^2 &= \left|\left|y - H_1s_1 - \sum\limits_{i=2}^{n}{H_is_i}\right|\right|^2 \\
	& = \left|\left|y - H_1s_1 - \begin{bmatrix} H_2 & \cdots & H_n \end{bmatrix} t_1 \right|\right|^2 \\
	&= (y - H_1 s_1)^{\dagger}(I_{n^2}-G_1 (G_1^{\dagger} G_1)^{-1} G_1^{\dagger})(y-H_1s_1) \\ &+ (\tilde{t}_1(s_1)-t_1)^{\dagger} G_1^{\dagger} G_1(\tilde{t}_1(s_1)-t_1), \nonumber
\end{align}
where $G_1 = \begin{bmatrix} H_2 & \cdots & H_n \end{bmatrix}$, and 
\begin{equation}
	\tilde{t}_1(s_1) = (G_1^{\dagger} G_1)^{-1} G_1 (y-H_1s_1). 
\end{equation} 
Here we are defining $\tilde{t}_1(s_1)$ to be an approximation of $t_1$ given $s_1$.  The approximation is essentially chosen to minimize the value of the quadratic form $v^\dagger G_1^\dagger G_1v$, which in turn should minimize the expression in (\ref{eqn:argmin}).  The success of this process will depend on the determinant and condition number of $G_1^\dagger G_1$ in the manner stated above.

Let $\mathfrak{Q}$ be a quantizer for the signaling alphabet $\mathcal{S}$.  Since $\tilde{t}_1(s_1)$ is not necessarily a constellation point, we now decode an estimate 
\begin{equation}
	\hat{t}_1(s_1) = \mathfrak{Q}(\tilde{t}_1(s_1)),
\end{equation}
and substitute this function into (\ref{eqn:argmin}), which now only depends on $s_1$ and reads
\begin{equation}
	\argmin\limits_{s_1 \in \mathcal{S}^n}\left|\left|y-H_1s_1-G_1\hat{t}_1(s_1)\right|\right|^2.
\end{equation}
We can thus compute an estimate $\hat{s}_1$ of $s_1$. Having estimated $\hat{s}_1$, we can go back to (\ref{eqn:argmin}) and replace $y$ with $y-H_1\hat{s}_1$. Using the same procedure and reindexing, we can solve for $s_2$. The algorithm is done after $n$ steps.  We present the algorithm in pseudocode below.

% The complexity blabla just checking how much I need to write for the algorithm to really be where it should be cause it's too big and makes everything look weird. Basically, I need to write so much that the algorithm starts on a new page. This is not yet enough. But maybe we add a table. That should do it, and we need it anyway. 
% \begin{center}
% 	\begin{tabular}{c||c|c}
% 	& Recursive & ML-decoding \\
% 	\hline \hline
%	Complexity & & \\
%	\hline
%	Smth else & & 
%	\end{tabular}
%\end{center} 

\begin{center}
	\begin{algorithm}[h!]
	\caption{Recursive Decoding of ST Codes}
		\begin{algorithmic}[1]
			\State \textbf{Input:} $Y, H$
			\State \textbf{Output:} $s_1,\ldots,s_n$
			\lskip
			\State \textbf{\underline{Step 1:}} 
				\For {$i = 1,\ldots,n$} 
					\State Compute $H_i = (I_n\otimes(H\rho(\omega_i)^t))\Gamma_i$ 
				\EndFor 
				\State Define $\mathcal{H} = \left\{H_1,\ldots,H_n\right\}$, $\mathcal{I} = \left\{1,\ldots,n\right\}$
				\State Compute $y = \vect(Y)$
			\lskip
			\State \textbf{\underline{Step 2:}} 
				\While{$\mathcal{H} \neq \emptyset$}
					\For{$i \in \mathcal{I}$} 
						\State Define $G_i = \begin{bmatrix} H_1 & \cdots & \hat{H}_i & \cdots & H_n \end{bmatrix}$ where $H_j \in \mathcal{H}$, $j \neq i$. 
					\EndFor
					\State Compute \begin{equation}i_1 = \argmax\limits_{i \in \mathcal{I}}\left\{\frac{\det(G_i^{\dagger}G_i)}{\kappa(G_i^{\dagger}G_i)}\right\}\end{equation}
					\State Update $\mathcal{H} = \mathcal{H}\backslash\left\{H_{i_1}\right\}$, $\mathcal{I} = \mathcal{I}\backslash\left\{i_1\right\}$			
				\lskip
				\State \textbf{\underline{Step 3:}} 
					\State Compute \begin{equation}\hat{s}_{i_1} = \argmin\limits_{s_{i_1} \in \mathcal{S}^n}{\left|\left|y- H_{i_1}s_{i_1} - G_{i_1}\mathfrak{Q}((G_{i_1}^{\dagger}G_{i_1})^{-1}G_{i_1}(y-H_{i_1}s_{i_1}))\right|\right|^2}\end{equation}
					\State Update $y = y - H_{i_1}\hat{s}_{i_1}$
					\State \textbf{goto Step 2}
				\EndWhile
		\end{algorithmic}	
	\end{algorithm}
\end{center}

\section{The $2\times 2$ Case}
\label{sec:2x2}

As a special case of the above scenario, set $n = 2$ and let $\LK$ be a degree $2$ real extension of number fields, that is $\LL = \K(\sqrt{d})$ for a squarefree $d \in \Z_{>1}$, and $\Cl_{\K} = 1$. Let $\OK$ and $\OL = \OK[\omega]$ be the ring of integers of $\K$ and $\LL$, respectively. Necessarily, the Galois group $\GLK = \langle \sigma \rangle$ of $\LK$ is cyclic of order $2$, where $\sigma$ is defined by $\sigma:\sqrt{d}\mapsto-\sqrt{d}$. 

Choose $\gamma \in \K^{\times}$ such that $\gamma \notin \Nm_{\LK}(\LL^\times)$ and $|\gamma|^2 = 1$, and define the quaternion algebra 
\begin{equation}
	\mathcal{Q} = (d,\gamma)_\K \cong \LL \oplus e\LL,
\end{equation} 
where $e^2 = \gamma$. For a shaping element $\alpha\in L^\times$, the ST code constructed from the transpose of the left regular representation of $\mathcal{Q}$ is a subset 
\begin{equation}
	\mathcal{X} \underset{\text{\tiny{finite}}}{\subset} \left\{\left. \begin{bmatrix}\alpha & \\ & \sigma(\alpha) \end{bmatrix}\begin{bmatrix} x_1 + x_2\omega  & x_3 + x_4\omega \\ \gamma(x_3 + x_4\sigma(\omega)) & x_1 + x_2\sigma(\omega) \end{bmatrix}\right| x_i \in \OK, 1 \le i \le 4 \right\}.
\end{equation}

Given the transmission model in (\ref{eqn:mimo}) for a channel matrix $H = \left[\begin{smallmatrix} h_{11} & h_{12} \\ h_{21} & h_{22} \end{smallmatrix}\right]$, vectorizing the received matrix results in the equivalent equation
\begin{equation}
	\vect(Y) = \vect(HX) + \vect(N) = H_1 s_1 + H_2 s_2 + \vect(N),
\end{equation} 
where $s_1 = [x_1\ x_2]^t$, $s_2 = [x_3\ x_4]^t$, and 

\begin{equation}
H_1 = \begin{bmatrix}
\alpha h_{11} & \sigma(\alpha)\gamma h_{12} \\
\sigma(\alpha)h_{12} & \alpha h_{11} \\
\alpha h_{21} & \sigma(\alpha)\gamma h_{22} \\
\sigma(\alpha) h_{22} & \alpha h_{21}
\end{bmatrix},
\quad
H_2 = \begin{bmatrix}
\alpha\tau h_{11} & \sigma(\alpha\tau)\gamma h_{12} \\
\sigma(\alpha\tau)h_{12} & \alpha\tau h_{11} \\
\alpha\tau h_{21} & \sigma(\alpha\tau)\gamma h_{22} \\
\sigma(\alpha\tau)h_{22} & \alpha\tau h_{21}
\end{bmatrix}.
\end{equation}
The relevant covariance matrices are
% The decoding algorithm proposed in \cite{sirinaunpiboon} attempts to decode $s_1$ and $s_2$ successively, that is decodes two groups involving two variables each, and decides which to decode first based on certain parameters of the covariance matrices
\begin{equation}
	H_1^\dagger H_1 = \begin{bmatrix} a & b \\ b^\ast & a \end{bmatrix}\quad \text{and}\quad H_2^{\dagger} H_2 = \begin{bmatrix} a_\omega & \Nm_{\LK}(\omega)b \\ (\Nm_{\LK}(\omega)b)^\ast & a_\omega \end{bmatrix},
\end{equation}
where the superscript $\ast$ denotes complex conjugation, and  
\begin{align}
	a &= ||\alpha H^{(1)}||^2 + ||\sigma(\alpha) H^{(2)}||^2, \\
	a_\omega &= ||\alpha\omega H^{(1)}||^2 + ||\sigma(\alpha\omega) H^{(2)}||^2, \\ 
	b &= \langle \alpha H^{(1)},\sigma(\alpha) H^{(2)} \rangle + \langle \sigma(\alpha)\gamma H^{(2)}, \alpha H^{(1)} \rangle.
\end{align}
where $H^{(i)}$ is the $i^{\text{th}}$ column of $H$.  These covariance matrices exhibit a very simple form due to the assumptions that $\LL$ is a real extension of $\K$ and $|\gamma|^2 = 1$. 

The determinants and condition numbers can easily be computed to be 
\begin{center}
	\begin{tabular}{l|cccc}
	& & $H_1^{\dagger}H_1$ & & $H_2^{\dagger}H_2$ \\
	\hline
	& & & &\\
	$\det(\cdot)$ & & $a^2-|b|^2$ & & $a_{\omega}^2-|\Nm_{\LK}(\omega)b|^2$ \\
	& & \\
	$\kappa(\cdot)$ & & $\frac{a+|b|}{a-|b|}$ & & $\frac{a_{\omega}+|\Nm_{\LK}(\omega)b|}{a_{\omega}-|\Nm_{\LK}(\omega)b|}$ 
	\end{tabular}
\end{center}
Thus, for example, if 
\begin{equation}
\frac{\det(H_2^{\dagger}H_2)}{\kappa(H_2^{\dagger}H_2)} \ge \frac{\det(H_1^{\dagger}H_1)}{\kappa(H_1^{\dagger}H_1)}
\end{equation}
then we define the following function of $s\in\mathcal{S}^2$
\begin{equation}
\tilde{s}_2(s) = (H_2^\dagger H_2)^{-1}H_2(y-H_1s)
\end{equation}
and subsequently estimate $s_1$ and $s_2$ by computing
\begin{align}
\hat{s}_1 = \argmin_{s\in\mathcal{S}^2}||y - H_1s - H_2\hat{s}_2(s)||^2, \ \
\hat{s}_2 = \argmin_{s\in\mathcal{S}^2}||y-H_1\hat{s}_1 - H_2s||
\end{align}

For the Golden code \cite{belfiore}, which is the code of interest in \cite{sirinaunpiboon}, the quaternion algebra used for code construction is $\mathcal{Q}_{G} = (5,i)_{\Q(i)}$ with $\omega = \frac{1+\sqrt{5}}{2}$. In this special case it is easy to see that the covariance matrix with the larger determinant also has the smaller condition number. Thus one can reduce to the decision metric to simply computing $\argmax_{i = 1,2} \{\det(H_i^\dagger H_i)\}$.

\section{The $4\times 4$ Perfect Code}
\label{sec:sims}
In this section we illustrate the decomposition of Section~\ref{sec:stc} for the $4\times 4$ perfect code. Let $\LK = \Q(i,\theta)/\Q(i)$, where $\theta = \zeta_{15}+\zeta_{15}^{-1} = 2\cos\left(2\pi/15\right)$. $\LK$ is a cyclic extension of degree $4$, and its Galois group is $\GLK = \langle \sigma \rangle$, where $\sigma: \zeta_{15}+\zeta_{15}^{-1} \mapsto \zeta_{15}^2+\zeta_{15}^{-2}$.  Consider the central simple algebra $\mathcal{A} = \bigoplus\limits_{j=0}^{3}{e_j \LL}$ as defined in Section~\ref{sec:stc}.  If $x\in \mathcal{A}$ and $\rho$ is the left regular representation, a codeword is of the form $X = \rho(x)^t$ and the corresponding decomposition of $\vect(HX)$ is
\begin{equation}
\vect(HX) = \sum_{i = 1}^4 (I_4\otimes (H\rho(\omega_i)))\Gamma s_i
\end{equation}
where the $\omega_i$ are given explicitly by $\omega_1 = (1-3i) + i\theta^2$, $\omega_2 = (1-3i)\theta + i\theta^3$, $\omega_3 = -i + (-3+4i)\theta + (1-i)\theta^3$, and $\omega_4 = (-1+i) - 3\theta + \theta^2 + \theta^3$
and the matrix $\Gamma = [\Gamma_1^t \cdots \Gamma_4^t]^t$ is given by
\begin{equation}
\Gamma_1 = \begin{bmatrix}
1 & 0 & 0 & 0 \\
0 & 0 & 0 & \gamma \\
0 & 0 & \gamma & 0 \\
0 & \gamma & 0 & 0
\end{bmatrix}, \
\Gamma_2 = \begin{bmatrix}
0 & 1 & 0 & 0 \\
1 & 0 & 0 & 0 \\
0 & 0 & 0 & \gamma \\
0 & 0 & \gamma & 0
\end{bmatrix}, \
\Gamma_3 = \begin{bmatrix}
0 & 0 & 1 & 0 \\
0 & 1 & 0 & 0 \\
1 & 0 & 0 & 0 \\
0 & 0 & 0 & \gamma
\end{bmatrix}, \
\Gamma_4 = \begin{bmatrix}
0 & 0 & 0 & 1 \\
0 & 0 & 1 & 0 \\
0 & 1 & 0 & 0 \\
1 & 0 & 0 & 0
\end{bmatrix}
\end{equation}
% We should remark that the $\omega_i$ are a $\Z[i]$-basis of the ideal $(\omega_1)$ of $\mathcal{O}_L$. 
\begin{figure}[h!]
\centering
\includegraphics[width=.49\textwidth]{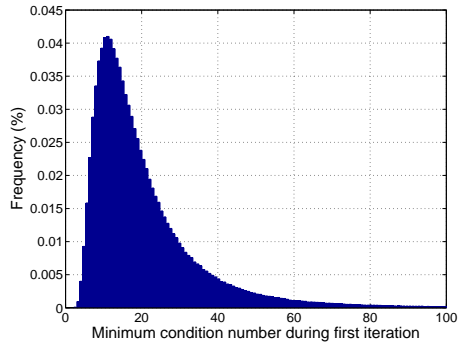}
\caption{The empirical distribution of $\min_i\{\kappa(G_i^\dagger G_i)\}$ during the first round of recursive decoding.}
\end{figure}

\vspace{-.75cm}
In Fig.\ 1 we plot the empirical distribution of $\min_i\{\kappa(G_i^\dagger G_i)\}$ averaged over $10^6$ channel matrices.  As in \cite{sirinaunpiboon}, we notice that the minimal condition number is quite well-behaved, hence it is highly likely that one of the four linear systems in question is well-conditioned and thus we can expect decoding to be successful for a high percentage of channels.  In contrast with \cite{sirinaunpiboon}, it is no longer true that the index minimizing the condition number is the one maximizing the determinant.  While we are not deciding based solely on the condition number, the index minimizing the condition number is highly correlated with the one maximizing the ratio, having correlation coefficient $\approx 0.6470$.  

\begin{figure}[h!]
\centering
\includegraphics[width=.49\textwidth]{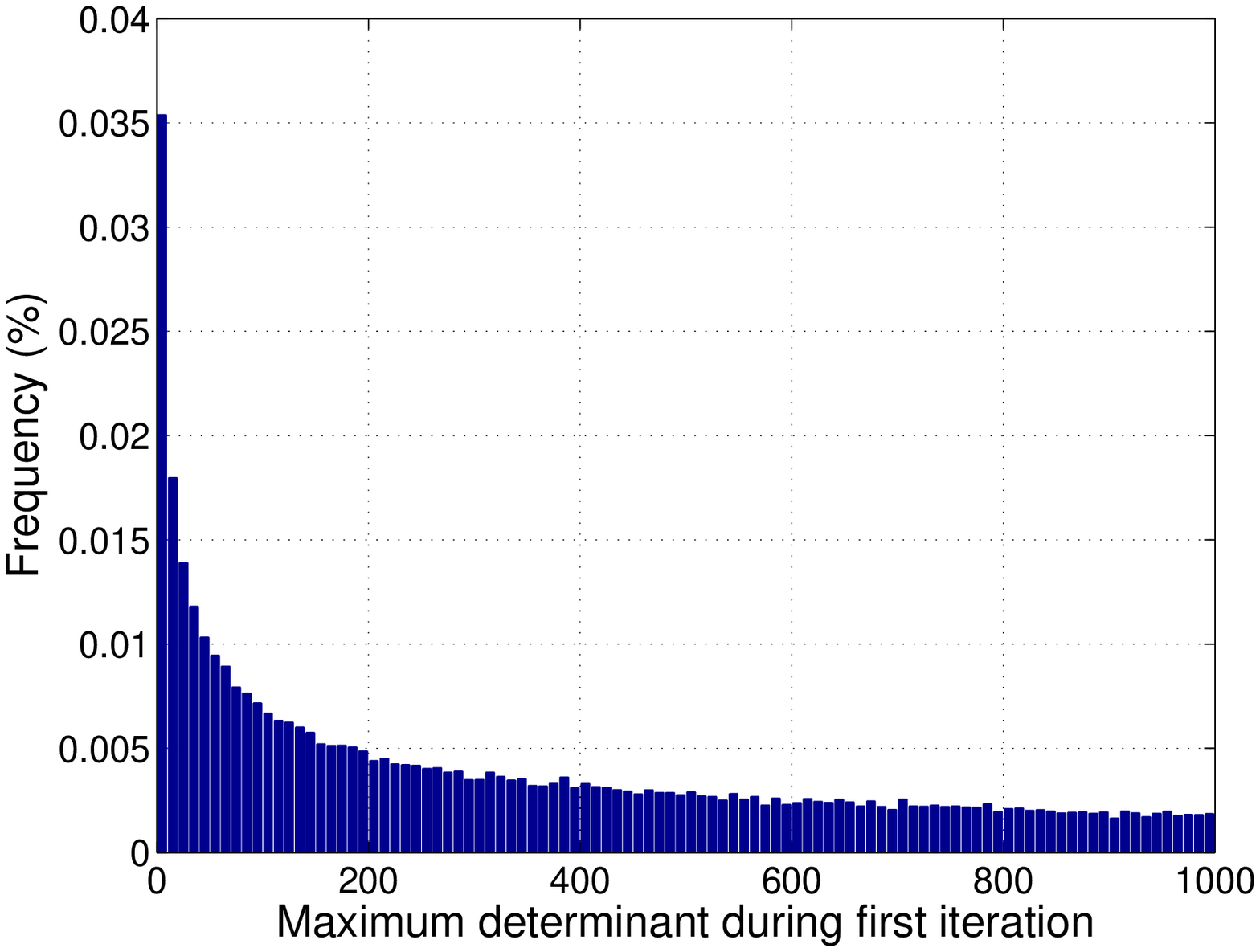}
\includegraphics[width=.49\textwidth]{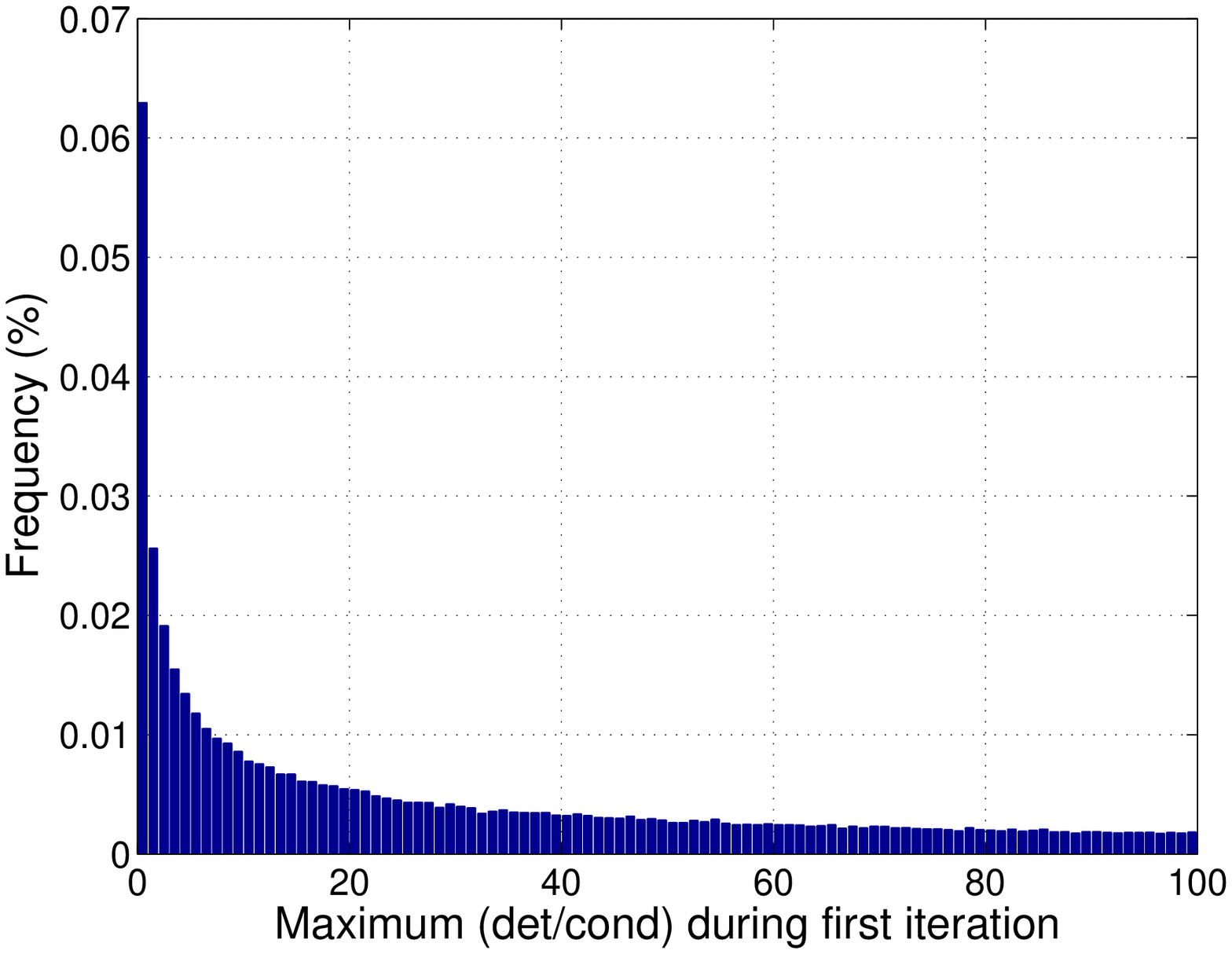}
\caption{The empirical distributions of $\max_i\{\det(G_i^\dagger G_i)\}$ and $\max_i \{\det(G_i^\dagger G_i)/\kappa(G_i^\dagger G_i)\}$, respectively, during the first round of recursive decoding.}
\end{figure}

In Fig.\ 2 we compare the empirical distributions of $\max_i\{\det(G_i^\dagger G_i)\}$ and $\max_i\{\det(G_i^\dagger G_i)/\kappa(G_i^\dagger G_i)\}$.  The similarity of the histograms is a symptom of the high correlation coefficient of $\approx 0.9095$ between the $i$ maximizing the determinant, and that which maximizes the ratio.  Thus the decision metric in the algorithm essentially amounts to selecting the index maximizing the determinant, with the condition number selecting the better-conditioned linear system among several with similarly-sized determinants.

\section{Conclusions and Future Work}
\label{sec:con}
In this article, we have proposed a recursive decoding method for algebraic space--time codes of arbitrary size, which has complexity $O(|\mathcal{S}|^n)$ compared to the worst-case complexity of $O(|\mathcal{S}|^{n^2})$ of ML-decoding.  The algorithm is based on a decomposition of the representation of the corresponding central simple algebra into a sum of $n$ matrices, each of which encodes $n$ information symbols.  These groups of symbols are then decoded recursively, based on certain properties of equivalent channel sub-matrices.  As basic examples, we have studied the relevant algebraic decompositions for $2\times 2$ codes and the $4\times 4$ perfect code, as well as provided a first statistical understanding of the algorithm for the latter.  

The most immediate task at hand is that of experimental error rates for various algebraic space-time codes, comparing our algorithm with ML-decoding and with the algorithms proposed in \cite{sirinaunpiboon,howard,amani}.  This simulation results are forthcoming and will appear in a later version of this paper.  One promising avenue of future work is comparing and possibly combining this approach with that of fast-decodability.  The two approaches appear mutually exclusive, and thus could be combined to further reduce decoding complexity.

\nocite{*}
\bibliographystyle{splncs}
\bibliography{refs}

\end{document}